\documentclass[usenatbib]{mn2e}
\usepackage{color}
\usepackage{ulem}

\usepackage{epsf}
\usepackage{graphicx}
\def\gsim { \lower .75ex \hbox{$\sim$} \llap{\raise .27ex \hbox{$>$}} }
\def\lsim { \lower .75ex \hbox{$\sim$} \llap{\raise .27ex \hbox{$<$}} }

\begin{document}

\title[The total density profile of DM halos]{The total density profile of DM 
halos fitted from strong lensing} 
\author[Wang et al.]{
\parbox{\textwidth}{Lin Wang$^{1,2}$\thanks{E-mail: wl010@bao.ac.cn}, 
Da-Ming Chen$^{1,2}$}
\vspace*{4pt} \\
$^1$National Astronomical Observatories, Chinese Academy of Sciences, 20A Datun 
Road, Chaoyang District, Beijing 100012, China; \\
$^2$School of Astronomy and Space Science, University of Chinese Academy of 
Sciences, Beijing 100049, China }
\maketitle

\begin{abstract}
In cosmological N-body simulations, the baryon effects on the cold dark matter 
(CDM) halos can be used to solve the small scale problems in $\Lambda$CDM 
cosmology, such as cusp-core problem and missing satellites problem. It turns 
out that the resultant total density profiles (baryons plus CDM), for halos 
with mass ranges from dwarf galaxies to galaxy clusters, can match the 
observations of the rotation curves better than NFW profile. In our previous 
work, however, we found that such density profiles fail to match the most 
recent strong gravitational lensing observations. In this paper, we do the 
converse:  we fit the most recent strong lensing observations  with the 
predicted lensing probabilities based on the so-called $(\alpha,\beta,\gamma)$ 
double power-law profile, and use the best-fit parameters ($\alpha=3.04, 
\beta=1.39, \gamma=1.88$) to calculate the rotation curves. We find that, at 
outer parts for a typical galaxy, the rotation curve calculated with our fitted 
density profile is much lower than observations and those based on 
simulations, including the NFW profile. This again verifies and strengthen the 
conclusions in our previous works: in $\Lambda$CDM paradigm, it is difficult to 
reconcile the contradictions between the observations for rotation curves and 
strong gravitational lensing.
\end{abstract}
\begin{keywords}
gravitational lensing: strong---galaxies: haloes---cosmology: theory---dark matter.
\vspace*{-0.5 truecm}
\end{keywords}

\section{Introduction}
\label{sec:intro}
Most people believe that we are now entering a period of precise cosmology: an 
inflation follows immediately after the big bang, and about 13.7 billion years 
later, we are living in a flat universe suffused with cosmological constant 
($\Lambda$) and cosmic webs made up of cold dark matter (CDM) particles and 
baryons. In short, we have a standard $\Lambda$CDM cosmology. There is a wide 
range of observational successes supporting this $\Lambda$CDM scenario, from 
cosmic microwave background (CMB) to the Lyman-$\alpha$ forest, to galaxy 
clustering and to weak gravitational lensing. However, three of the key 
ingredients remains mystery: $\Lambda$, CDM, and the scalar field that drives 
inflation. There are no independent experimental evidences or strong theoretical 
justifications for them. Clearly, they are introduced so that the previously 
mentioned observational successes can be achieved based on general relativity. 

If we set the mystery aspects of $\Lambda$CDM cosmology aside, and enjoy our 
hard-won successes mainly on large scales, we still encounter some difficulties 
concerned with dark matter (DM) on small scales. 
After the universe enters matter-dominated era and nonlinear perturbations 
begin, the current computer simulations become the most robust tool to explore 
the formation and evolution of the large scale structure \citep{FW12}. In the 
standard, hierarchical, CDM paradigm of cosmological structure formation, galaxy 
formation begins with the gravitational collapse of
over dense regions into bound, virialized halos of CDM.  In this $\Lambda$CDM 
paradigm, halos form from purely collisionless CDM particles with primordial 
power spectrum of fluctuations predicted by inflationary model. Small halos are 
the first to form, and larger halos form subsequently by mergers of 
pre-existing halos and by accretion of diffuse dark matter that has never been 
part of a halo. While such purely CDM N-body simulations can reproduce the 
observed cosmic web as demonstrated by Sloan Digital Sky Survey (SDSS), two 
tensions with observations arise on small scales. First, simulations show that 
CDM collapse leads to cuspy halos whose central density profiles have the form 
$r^{-\gamma}$ with $\gamma\sim 1-1.5$, whereas observed galaxy rotation curves 
suggest constant density cores with $\gamma\sim 0$. This conflict is referred 
as ``cusp-core problem''. Second, simulations predict a large amount of 
subhalos formed by earlier collapses on smaller scales, but astronomers have 
observed much less number of satellite galaxies, which is known as the 
``missing satellite problem''.  Recent simulations strongly suggest some 
connections between the two problems, or even that the two problems have merged 
into one \citep{Weinberg15}. 

A main type of solutions to the small scale problems is including baryons as 
``astrophysical processes'' in originally collisionless CDM N-body simulations. 
This is natural in $\Lambda$CDM cosmology. On one hand, if we replace CDM with 
warm dark matter, the small scale problems can be resolved even without 
considering the influence of baryons (e.g.,\citealt{SGTF12}). However, this is 
not $\Lambda$CDM cosmology anymore. On the other hand, the direct observable 
parts of any bound systems are made up of baryons rather than DM, so simulations 
must include baryons if we want to reproduce the disk galaxies and elliptical 
galaxies appear in our telescopes. In fact, in a simplified picture 
(\citealt{WR78}), baryonic gas is 
initially well mixed with the DM particles, then participates in the 
gravitational collapse of DM and is heated by shocks to the virial temperature 
of the DM halos.  Bound in the potential wells of DM halos, baryonic gas 
proceed to cool radiatively due to bremsstrahlung, recombination and 
collisionally exited line emission \citep{FW12}. Just 
before gravitational collapse, angular momentum is transfered to the aspherical 
perturbations by gravitational torques exerted by neighboring clumps. This 
results in the formation of a gas disk, and once the disk has become 
centrifugally supported, stars can be formed. Furthermore, the spheroidal 
components of disk galaxies and elliptical galaxies form by major mergers or 
strong gravitational encounters of disk galaxies which can lead to the complete 
destructions of the pre-exising disks. 

In this paper, the major concern lies in the 
density profiles of CDM haloes based on which we can calculate rotation curves 
and strong lensing probabilities. The baryons have two opposite effects on the 
central mass density of DM halos. While stellar feedback and dynamical friction 
can induce expansion of the DM halo and produce a core (e.g., \citealt{SIR99}, 
\citealt{MCW06}, \citealt{MWC08}), the adiabatic contractions can steepen 
central density to the singular isothermal sphere (SIS) type 
\citep{BFFP86,GKKN04,GFS06}. Unfortunately, on one hand, a cored density 
profile would lead to an extremely low lensing rate compared with observations 
\citep{Chen05,LC09,CM10}; on the other hand, when SIS profile matches the 
observations of strong lensing by giant elliptical galaxies, it fails in 
fitting the inner parts of rotation curves which indicate central cores. The 
tension between the observations of rotation curves \citep{Katz16,Schaller15} 
and strong lensing remains for most recent high-resolution baryon+CDM 
simulations (\citealt{Wang17}, hereafter, paper I). One such simulations 
(\citealt{Cintio14}, DC14) introduce a mass-dependent density 
profile to describe the distribution of dark matter within galaxies, which 
takes into account the stellar-to-halo mass ratio ($M_\ast/M_{\rm halo}$) 
dependence of baryon effects on DM. Comparing with previous similar works, 
a real progress for DC14 model is that it gives a density profile with inner 
slope depending on halo mass. At low mass end, each halo display a central 
core, and for halos with increasing mass,  some astrophysical processes erase 
the central cores and steepen the inner slopes of the DM density profiles. 
Since the observations of rotation curves usually come from low mass halos (such 
as dwarf galaxies, low surface brightness galaxies), whereas strong lensing 
phenomena is dominated by giant ellipticals (hosted in high mass halos), one 
hopes that DC14 model may resolve the tension. This can happen if the inner 
slope $\gamma$ approaches 2 (SIS like) for halos that host giant ellipticals, 
which is required by strong lensing observations. However, the inner slope of 
DC14 model increases with halo mass only up to $\gamma \sim 1$ (NFW like) when 
halo mass reaches the high mass end ($\sim 10^{12}M_{\sun}$). Furthermore, the 
extrapolation of DC14 profile to halos with mass $>10^{12}M_{\sun}$ exhibits no 
monotonic increasing of inner slope, instead, it drops dramatically after that 
mass \citep{Wang17}. It thus comes as no surprise that the DC14 model generates 
the lensing probabilities that are much lower than both NFW and SIS models. 
Despite the failure for DC14 model to resolve the tension between rotation 
curves and strong lensing, it indeed reduces the tension. Recall that, before 
DC14, similar simulations give the cored density profiles which are independent 
of halo mass. Consequently, DC14 produces obviously higher lensing rates 
than cored isothermal sphere model \citep{Wang17}. Therefore, DC14 stands for a 
right direction for simulations which may finally resolve the tension.

Another important example is the investigation for the internal 
structure and density profiles of halos of mass $10^{10}-10^{14}M_{\sun}$ in 
the Evolution and Assembly of Galaxies and their Environment (EAGLE) 
simulations (\citealt{Schaller15}, 
Schaller15). In this mass range the total density profile is similar to NFW in 
the inner and outer parts, but has a slope of $-2$ at some radius $r_{\rm i}\sim 
2.27$kpc, near the centers of halos. Schaller15 profile has no core, however, 
the rotation curves are in excellent agreement with observational data 
\citep{Reyes11}. In paper I, we found that Schaller15 
profile predicts too many lensing probabilities compared with the observations 
of the Sloan Digital Sky Survey Quasar Lens Search (SQLS12, \citealt{Inada12}). 
This can be explained by the fact that, for Schaller15 halos, 
the central regions of halos with mass  $\gsim 10^{12}M_{\sun}$ are 
dominated by the stellar component \citep{Schaller15}. The presence of these 
baryons causes a contraction of the halos and thus enhances the density of DM 
in this regions. The over predictions of lensing probabilities for Schaller15 
profile also means a failure to resolve the tension between rotation curves and 
strong lensing.

It is now well established that, whatever the manners the baryon effects are 
included in the collisionless CDM N-body cosmological simulations, if the 
resultant density profiles can match the observations of rotation curves, they 
cannot simultaneously predict the observations of strong gravitational lensing 
(under- or over-predict). And for the case of typical galaxies, the reverse is 
also true, namely, the SIS profile preferred by strong lensing cannot be 
supported by the observations of rotation curves near the centers of galaxies. 
It is unclear whether or not the reverse is true for extremely large galaxies 
and clusters of galaxies, which are expected to have generated the large image 
separations in the most recent strong lensing sample SQLS12.  It is well known 
that, before the release of SQLS12, the SIS+NFW is a standard model to describe 
the well-defined statistical sample for strong lensing observational data. 
However, we found that this standard model fails in describing SQLS12 (see 
paper I). According to the standard model, SIS and NFW profiles are used to 
describe the giant elliptical galaxies and clusters of galaxies respectively, 
the transition in mass from galaxies to clusters occurs at $M_{\rm halo}\sim 
10^{13}M_{\sun}$. We have noticed in our calculations that, simply shifting the 
transition mass from $M_{\rm halo}\sim 10^{13}M_{\sun}$ to, e.g., 
$10^{13.5}M_{\sun}$ cannot improve the matches. Thus the failure for standard 
model to describe the new sample SQLS12 may imply the failure for NFW to be  an 
appropriate model for clusters of galaxies adopted to predict lensing 
probabilities\citep{Giocoli16}. If this is the case, however, we have a paradox. 
We know that every statistical sample consists of individual lensing systems, 
each of which is investigated and identified separately. Among them, each 
individual cluster of galaxies is always modeled by NFW profile or triaxial form 
of NFW, and if needed, some substructures can be added only as perturbations. 
So, is it reasonable that the individuals can largely be modeled as NFW but 
their statistical sample as a whole cannot? 

Now that the results of current simulations, such as  DC14 and Schaller15 
profiles, are in good agreement with the observations of rotation curves but 
failure in predicting the strong lensing observations of SQLS12, it would be 
interesting to examine the converse: what a density profile, if exists, that 
matches the observations of SQLS12 predicts for rotation curves? To this end, 
in this paper, we further investigate the tension between the observations of 
rotation curves and strong lensing as follows. We first fit  the predicted 
lensing probabilities based on the so-called $(\alpha,\beta,\gamma)$ double 
power-law profile directly to the SQLS12 sample, and then use the best-fit 
parameters $(\alpha,\beta,\gamma)$ for the profile to calculate rotation curves. 
This method can circumvent the paradox, in the sense that we just employ the 
well-defined sample to derive an empirical formula for the density profile, 
regardless of the density profile models used for each individual lensing system 
of the sample.

This paper is organized as follows: in Section~\ref{sec:fit} we fit 
$(\alpha,\beta,\gamma)$ 
double power-law model to the most recent strong lensing sample SQLS12 to 
obtain a density profile of DM haloes. In Section~\ref{sec:rc}, the rotation 
curves calculated based on the fitted density profile are compared with 
observations and other models. The discussions and
conclusions are presented in Section~\ref{sec:conc}.

\section{Fitting the density profile}
\label{sec:fit}
The predicted lensing probabilities are determined by the assumed density 
profile of lensing objects. Therefore, in order to obtain an empirical formula 
for a density profile from strong lensing sample, one should first assume a 
functional form of the density profile with some free parameters; then fit the 
predicted lensing probabilities (based on the assumed density profile) to that 
of SQLS12 sample to get the values of the parameters.

We employ the so-called $(\alpha,\beta,\gamma)$ 
double power-law model as the assumed density profile for lenses
\par
\begin{equation}
{\rho(r)}= \frac{\rho_{\rm s}}{\left(\frac{r}{r_{\rm 
			s}}\right)^{\gamma}\left[1+\left(\frac{r}{r_{\rm 
			s}}\right)^{\alpha}\right]^{(\beta-\gamma)/\alpha}},
\label{eq:fitting}
\end{equation}
where $\rho_\mathrm{s}$ is the scale density and $r_\mathrm{s}$ the scale 
radius. In this paper, we treat $(\alpha,\beta,\gamma)$ as free constant 
parameters to be determined from the fitting, in contrast to DC14 model, 
in which they are halo mass dependent.

The corresponding surface mass density  is
\begin{eqnarray}
    \Sigma(x) = 2\rho_s r_s V(x)
    \label{sur}
\end{eqnarray}
where
\[
V(x)=\int_0^{\infty}\left(x^2+z^2\right)^{-\gamma/2}
       \left[\left(x^2+z^2\right)^{\alpha/2}+1\right]^{(\gamma-\beta)/\alpha}dz,
\]
and $x=|\vec{x}|$, $\vec{x} = \vec{\xi}/r_s$, $\vec{\xi}$ is the position
vector in the lens plane.  We thus obtain the lensing equation 
\begin{eqnarray}
    y = x - \mu_s {g(x)\over x}\,,
    \label{lens1}
\end{eqnarray}
where $y = |\vec{y}|$, $\vec{\eta} = \vec{y}\, r_s D_\mathrm{S}/D_\mathrm{L}$ 
is the position 
vector in the source plane, and
\begin{eqnarray}
    g(x) \equiv \int_0^x u V(u) du,
    \label{gx}
\end{eqnarray}
and
\begin{eqnarray}
    \mu_s \equiv {4\rho_s r_s\over \Sigma_{\rm cr}}\,,
    \label{mus}
\end{eqnarray}
where $\Sigma_\mathrm{cr}=(c^2/4\pi
G)(D_\mathrm{S}/D_\mathrm{L}D_\mathrm{LS})$ is
the critical surface mass density; $D_\mathrm{L}$, $D_\mathrm{S}$  and
$D_\mathrm{LS}$ are the angular diameter distances from the observer to the 
lens, to the source and from the lens to the source, respectively.

When the quasars of redshift $z_{\mathrm{s}}$ are lensed by foreground CDM 
halos of galaxy clusters and galaxies, the lensing probability with image 
separations larger than $\Delta\theta$ is \citep{schne}
\begin{eqnarray}
P(>\Delta\theta)&=&\int^{z_{\mathrm{s}}}_0\frac{dD^\mathrm{P}_{\mathrm{L}}(z)}
{dz}dz \nonumber \\
& &\times\int^{\infty}_0\bar{n}(M,z)\sigma(M,z)B(M,z)dM ,
\label{prob1}
\end{eqnarray}
where  $D^\mathrm{P}_{\mathrm{L}}(z)$ is the
proper distance from the observer to the lens located at redshift $z$. We make 
$z_{s}=1.56$ for statistical sample SQLS (\citealt{Inada12}). The physical 
number density $\bar{n}(M,z)$ of virialized DM halos of 
masses between $M$ and $M+dM$ is related to the comoving number density 
$n(M,z)$ by $\bar{n}(M,z)=n(M,z)(1+z)^3$. As in paper I, $n(M,z)$ was 
originally given by \cite{press74}, and we use the improved version given by
\cite{Sheth99}. The cross-section is
$\sigma(M, z)=\pi
y_{\mathrm{cr}}^2r_{\mathrm{s}}^2\vartheta(M-M_{\mathrm{min}})$,
where $y_\mathrm{cr}$ is the maximum value of $y$, the reduced position of a
source, such that when $y<y_\mathrm{cr}$ multiple images can occur;  
$\vartheta(x)$ is a step function, and $M_{\mathrm{min}}$ is the 
minimum value of the halo mass that can produce image separation 
$\Delta\theta$. Also, as in paper I, for the magnification 
bias $B(M,z)$, we adopt a simple model  (\citealt{li02}):
$B\approx 2.2A_{m}^{1.1}$, 
with $A_m=D_\mathrm{L}\Delta\theta/(r_sy_\mathrm{cr})$.

Cleanly, the free parameters $(\alpha,\beta,\gamma)$ have entered into 
$P(>\Delta\theta)$ via equations (\ref{eq:fitting}), (\ref{sur}) and 
cross-section $\sigma$, together with magnification bias $B$, whence we get the 
corresponding predicted number counts of lenses with image separations larger 
than $\Delta\theta$ for the sample SQLS12 as
\begin{equation}
 n(>\Delta\theta; \alpha,\beta,\gamma)=NP(>\Delta\theta),
 \label{n_counts}
\end{equation}
where $N=50836$ is the number of source quasars from which 26 lenses are 
selected for the sample SQLS12 (\citealt{Inada12}).

 We use least-square fit method. Choose a merit function that measures the 
agreement between the data and the model with a particular choice of parameters. 
The merit function is conventionally arranged so that small values represent 
close agreement. The parameters of the model are then adjusted to achieve a 
minimum in the merit function, yielding best-fit parameters. The adjustment 
process is thus a problem in minimization in many dimensions.  At last we get 
the best-fit parameters: $\alpha$=3.04, $\beta$=1.39 and $\gamma$=1.88. 
 
\begin{figure}
\begin{center}
\mbox{
\hspace{-0.7cm}
\resizebox{6.7cm}{!}{\includegraphics{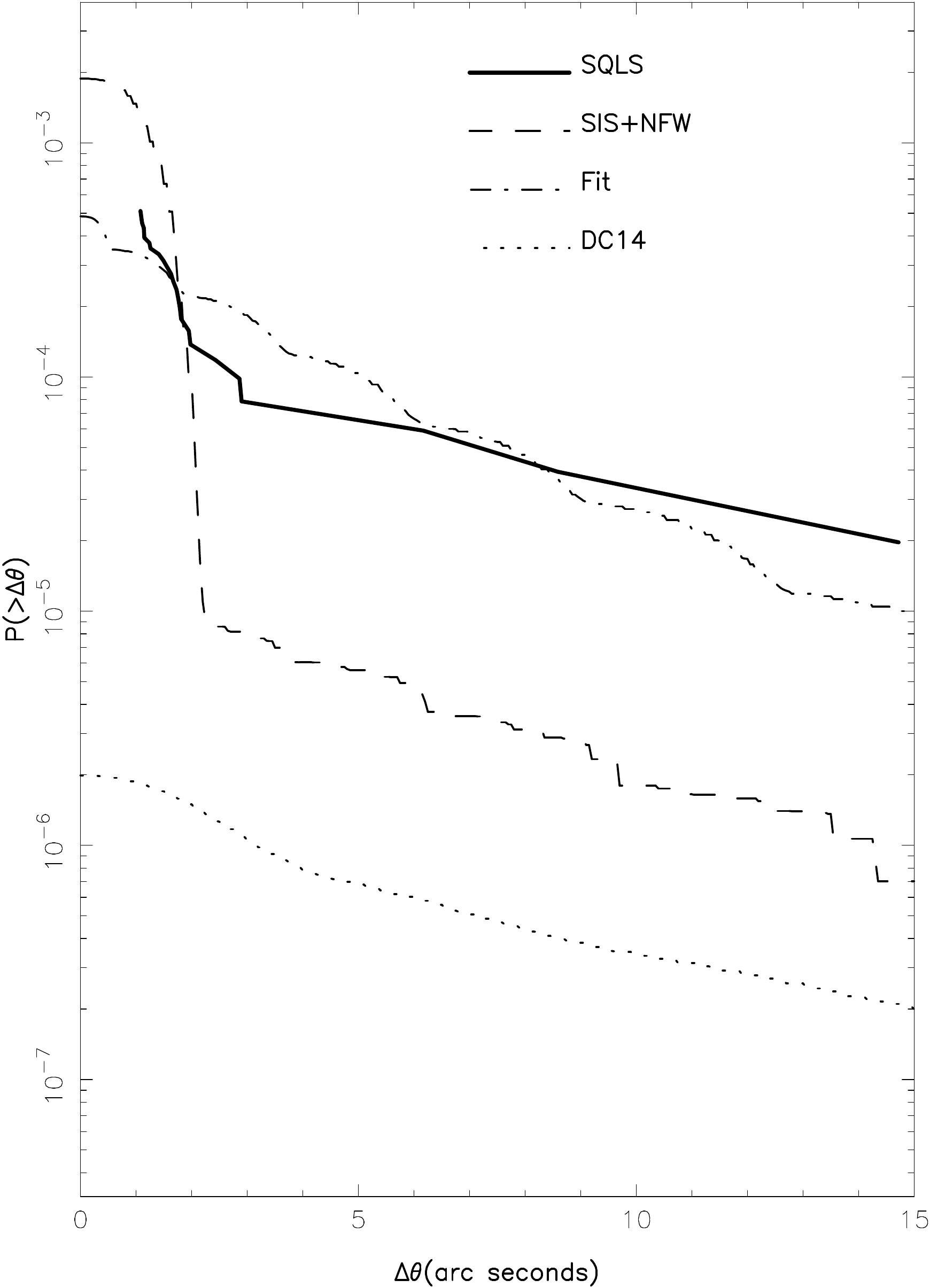}}
}
\caption{Lensing probabilities with image separations larger than 
$\Delta\theta$: observations for SQLS sample (thick histogram), and the
predictions for the models of SIS +NFW (dashed line), our fitting result 
(dot-dashed line) and DC14 (dotted line). }
\label{fig1}
\end{center}
\end{figure}

The predicted lensing probability for our fitting model (dot-dashed line) is 
displayed in Fig.\ref{fig1}, together with the observations of SQLS12 sample 
(thick histogram), the predictions for the models of SIS+NFW (dashed line) and 
DC14 (dotted line). The latter two lines are copied from paper I.

It turns out that our fitted one-population $(\alpha,\beta,\gamma)$ 
double power-law model can only roughly predict the strong lensing observations 
of SQLS12. Hence some two-population models, like SIS+NFW, are 
still preferred by better fittings. However, it suffices for the aim of this 
paper: We just need a density profile alternative to DC14 and Schaller15 models 
that can predict lensing observations of SQLS12 much better than the latter, 
which are also of one-population. 

\begin{figure}
\begin{center}
\mbox{
\hspace{-0.7cm}
\resizebox{6.7cm}{!}{\includegraphics{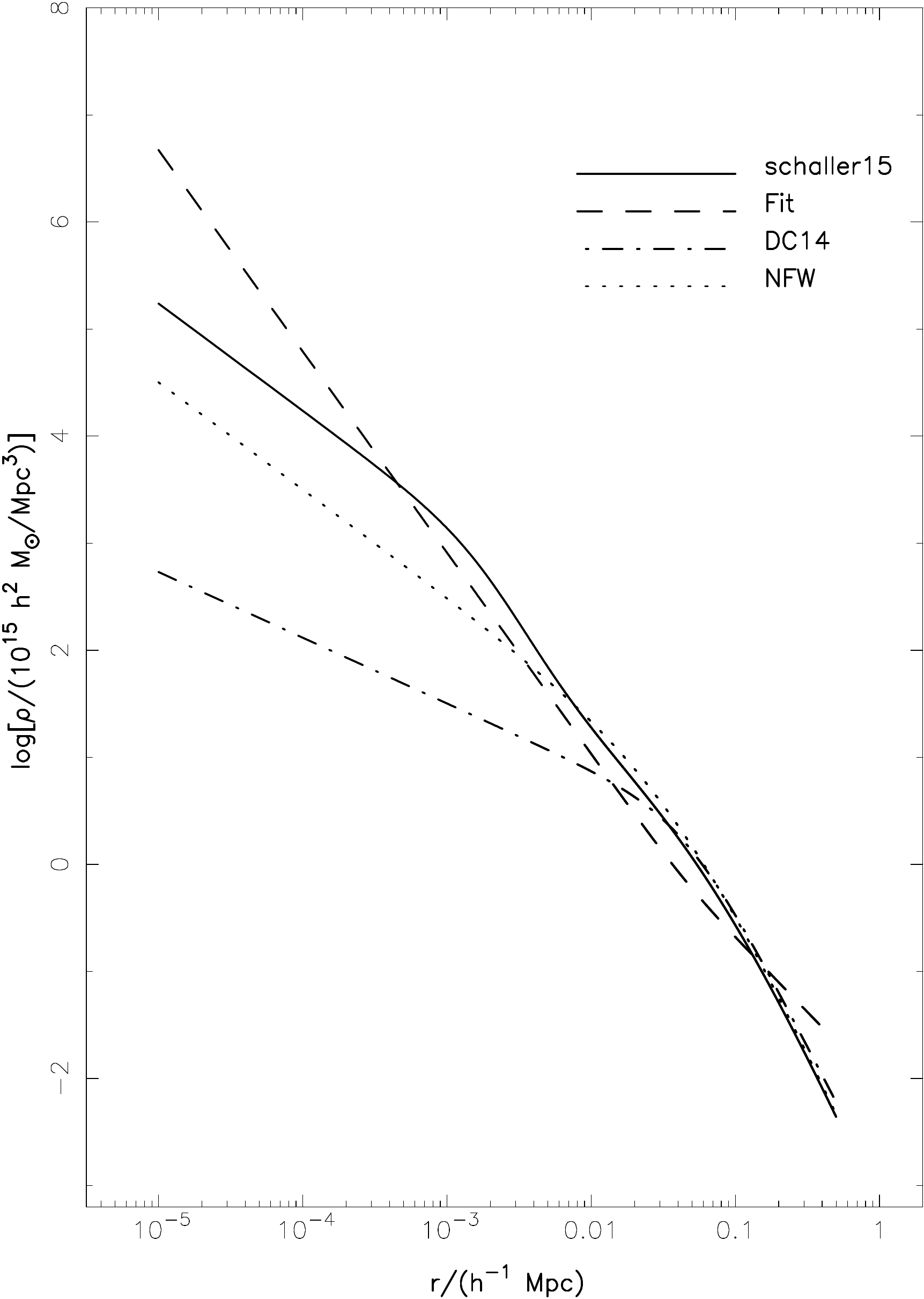}}
}
\caption{The density profiles for different models for a typical galaxy with 
halo mass $M_{\rm halo}=10^{13}M_{\sun}$: Schaller15 (solid 
line), our fit model (dashed line), DC14 (dot-dashed line) and NFW (dotted 
line).}
\label{fig2}
\end{center}
\end{figure}

Before investigating the rotation curves, it is helpful for us to compare some 
properties of the fitted density profile with other models, such as NFW, DC14 
and Schaller15.  Note that, for 
a given halo mass, the density profiles for different models can be plotted 
only when some additional parameters have been determined. For the 
$(\alpha,\beta,\gamma)$ double power-law model, such parameters are (see paper 
I)
\begin{equation}
\rho_\mathrm{s}=\rho_\mathrm{crit}\left[\Omega_\mathrm{m}(1+z)^3
+\Omega_{\Lambda}\right]\frac{200}{3}\frac{c_1^3}{f(c_1)},
\label{rhos}
\end{equation}

\begin{equation}
r_\mathrm{s}=\frac{1.626}{c_1}\frac{M_{15}^{1/3}}
{\left[\Omega_\mathrm{m}
(1+z)^3+\Omega_{\Lambda}\right]^{1/3}}h^{-1}\mathrm{Mpc}.
\label{rs}
\end{equation}
where $\rho_{crit}$ is the present value of the critical mass density of the 
universe, and $M_{15}$ is the reduced mass of a halo defined as
$M_{15}=M_\mathrm{halo}/(10^{15}\mathrm{h}^{-1}M_{\sun})$. The 
concentration parameter $c_1$ is approximately a constant and we have adopted 
$c_1\simeq 9$ (but see \citealt{Gao08}). And 
\begin{equation}
f(c_1)=\int^{c_1}_0\frac{x^2dx}{x^{\gamma}(1+x^{\alpha})^{(\beta-\gamma)/\alpha
}
}.
\end{equation}
Therefore, $\rho_{\rm s}$ and $r_{\rm s}$ are redshift-dependent, we choose 
$z=0.45$ for a typical lens. Clearly, DC14 and NFW models have the same 
functional forms of $\rho_{\rm s}$ and $r_{\rm s}$ as equations (\ref{rhos}) 
and (\ref{rs}), they differ only in different values of 
$(\alpha,\beta,\gamma)$. For DC14 these parameters are halo mass dependent, and 
for NFW, $(\alpha,\beta,\gamma)=(1,3,1)$. For Schaller15 profile,
\begin{equation}
\frac{\rho(r)}{\rho_{\rm crit}}= \frac{\delta_{\rm c}}{\left(\frac{r}{r_{\rm 
s}}\right)\left(1+\frac{r}{r_{\rm 
s}}\right)^{2}}+\frac{\delta_{\rm i}}{\left(\frac{r}{r_{\rm 
i}}\right)\left[1+\left(\frac{r}{r_{\rm 
i}}\right)^{2}\right]},
\label{Schaller}
\end{equation}
where the parameters $\delta_{\rm c}$, $\delta_{\rm i}$ and $r_{\rm s}$ are all 
fitted to halo mass with the data given by \citet{Schaller15}, and $r_{\rm 
i}=1.7{\rm h}^{-1}$kpc for all halo masses (see paper I).

\begin{figure}
\begin{center}
\mbox{
\hspace{-0.7cm}
\resizebox{6.7cm}{!}{\includegraphics{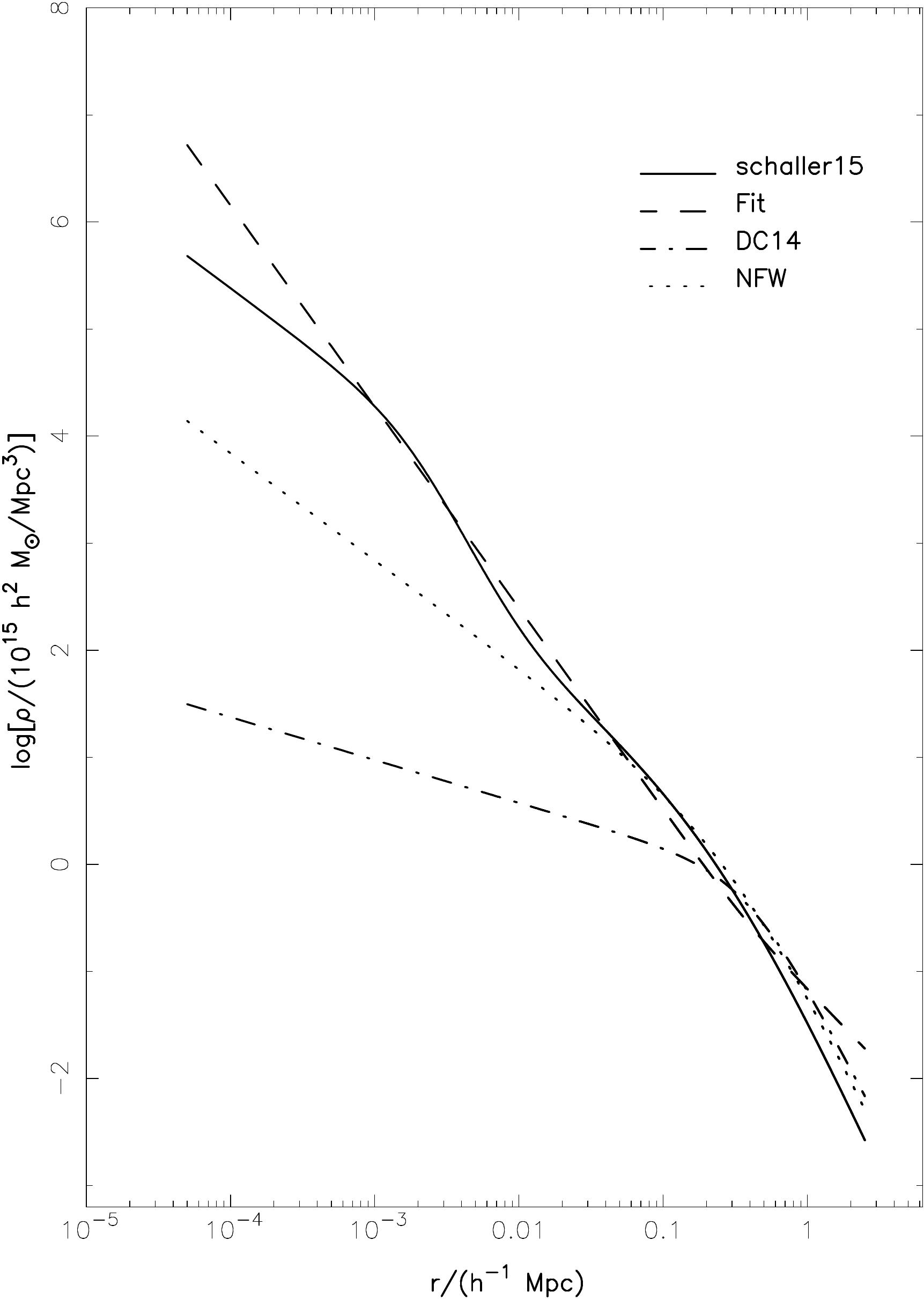}}
}
\caption{The density profiles for different models for a typical cluster of 
galaxies with halo mass $M_{\rm halo}=10^{15}M_{\sun}$:  
Schaller15 (solid line), our fit model (dashed line), DC14 (dot-dashed line) 
and NFW (dotted line).}
\label{fig3}
\end{center}
\end{figure}

The density profiles are presented for different models in Fig.\ref{fig2} and 
Fig.\ref{fig3} for the cases when a 
typical galaxy has halo mass of $M_{\rm halo}=10^{13}M_{\sun}$ and a typical 
cluster of galaxies has halo mass of $M_{\rm halo}=10^{15}M_{\sun}$, 
respectively. For a double power-law model, $\gamma$ and $\beta$ are the inner 
and outer slope respectively, and $\alpha$ describes the transition between the 
two. 

For a typical galaxy (Fig.\ref{fig2}), the slope of our fitted density 
profile ($\gamma=1.88$) is larger than other models in the innermost regions. 
This value is close to SIS model ($\gamma=2$), which is required to explain the 
observations for galactic haloes as lenses. The slope approaches 1.39 at outer 
regions, which is obviously lower than SIS ($\sim 2$) and NFW ($\sim 3$). 
Notice that, for Schaller15 profile, the slope has the value of 1 in the 
innermost regions, 2 at the radius $r=r_{\rm i}=1.7{\rm h}^{-1}$kpc (very near 
the center), and approaches to 3 in the outer regions (NFW like). This can 
explain the over-predictions to SQLS12 for Schaller15 model in terms of slopes. 
As for DC14 profile, in most regions starting from the center, the slope is 
flatter than NFW, and it reaches 3 (NFW like) only in very outer regions. This 
can explain  why DC14 model predict too few lenses.  

For a typical cluster of galaxies (Fig.\ref{fig3}), we find that our fitted 
density profile has almost the same slope as of Schaller15 in the regions 
between $r_{\rm i}\sim$kpc and $\sim$500kpc, whereas it is steeper than 
Schaller15 in the innermost regions ($< r_{\rm i}$), and flatter in outer 
regions ($\gsim 500$kpc). Therefore, the overpredictions of Schaller15 model 
(Fig.\ref{fig1}) suggest that the profiles with flatter outer slopes (flatter 
than NFW) are preferred.

\begin{figure}
\begin{center}
\mbox{
\hspace{-0.7cm}
\resizebox{6.7cm}{!}{\includegraphics{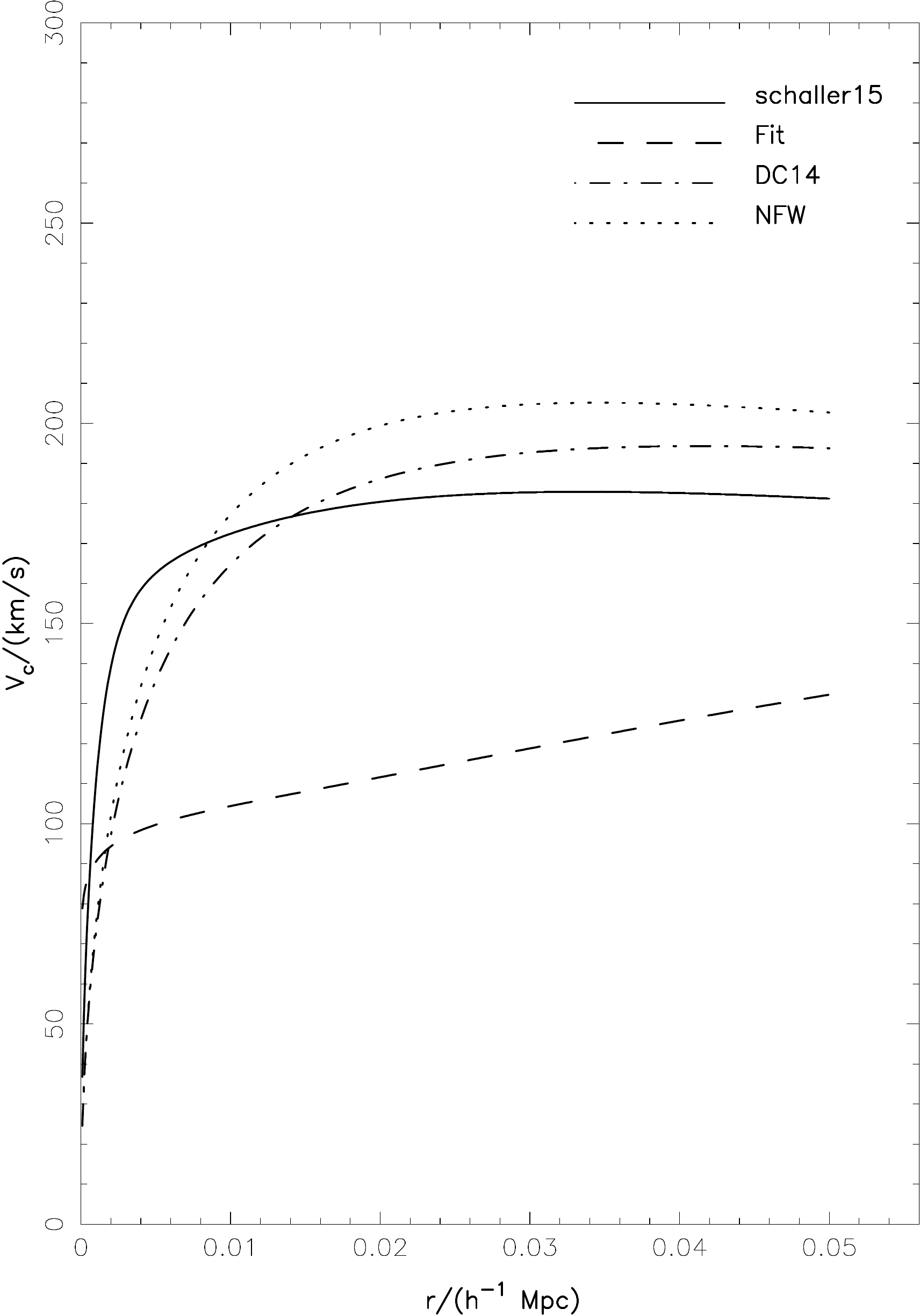}}
}
\caption{Rotation curves for different models: Schaller15 (solid 
line), our fit model (dashed line), DC14 (dot-dashed line) and NFW (dotted 
line).} 
\label{fig4}
\end{center}
\end{figure}

\section{rotation curves}
\label{sec:rc}
The observations of rotation curves provide us independent evidences for or 
against a density profile model. While simulations suggest some density 
profiles that are in agreement with the observations of rotation curves, 
e.g., Schaller15, we find that they cannot predict the most recent 
observations of strong gravitational lensing. It is thus very interesting to 
investigate the behavior of rotation curves for the density profile fitted from 
the observations of strong lensing in previous section. For a given 
spherically symmetric  density profile $\rho(r)$, the Poisson's equation 
\begin{equation}
\frac{1}{r^2}\frac{\partial}{\partial  r}\left( r^2 \frac{\partial 
\Phi}{\partial r}\right)=4 \pi G \rho
\end{equation}
leads to the circular velocity
\begin{equation}
{V_{c}}^2=r \frac{\mathrm{d}\phi}{\mathrm{d}r}=\frac{4\pi G}{r}\int_0^r 
\rho(r')r'^2dr', 
\label{eq:velocity}
\end{equation}
where we have required $\lim_{r\rightarrow 0} V(r)\rightarrow 0$.

Assuming a typical disk galaxy with halo mass $\sim 10^{12}M_{\sun}$, 
the rotation curves predicted by different density profiles can be obtained by 
numerical integrating via equation (\ref{eq:velocity}). The results are 
displayed in Fig.\ref{fig4} for the density profiles of our fit model 
(dashed), Schaller15 (solid), DC14 (dot-dashed) and NFW (dotted), respectively. 
It can be easily recognized that, in the flat part, the rotation curve of our 
fit model is much lower than that of other three models, which are around 
$200$km/s, a typical observed value for a halo with mass of $\sim 
10^{12}M_{\sun}$. Therefore, our fitted density profile from strong lensing 
observations is ruled out by the observations of rotation curves.

\begin{figure}
\begin{center}
\mbox{
\hspace{-0.7cm}
\resizebox{6.7cm}{!}{\includegraphics{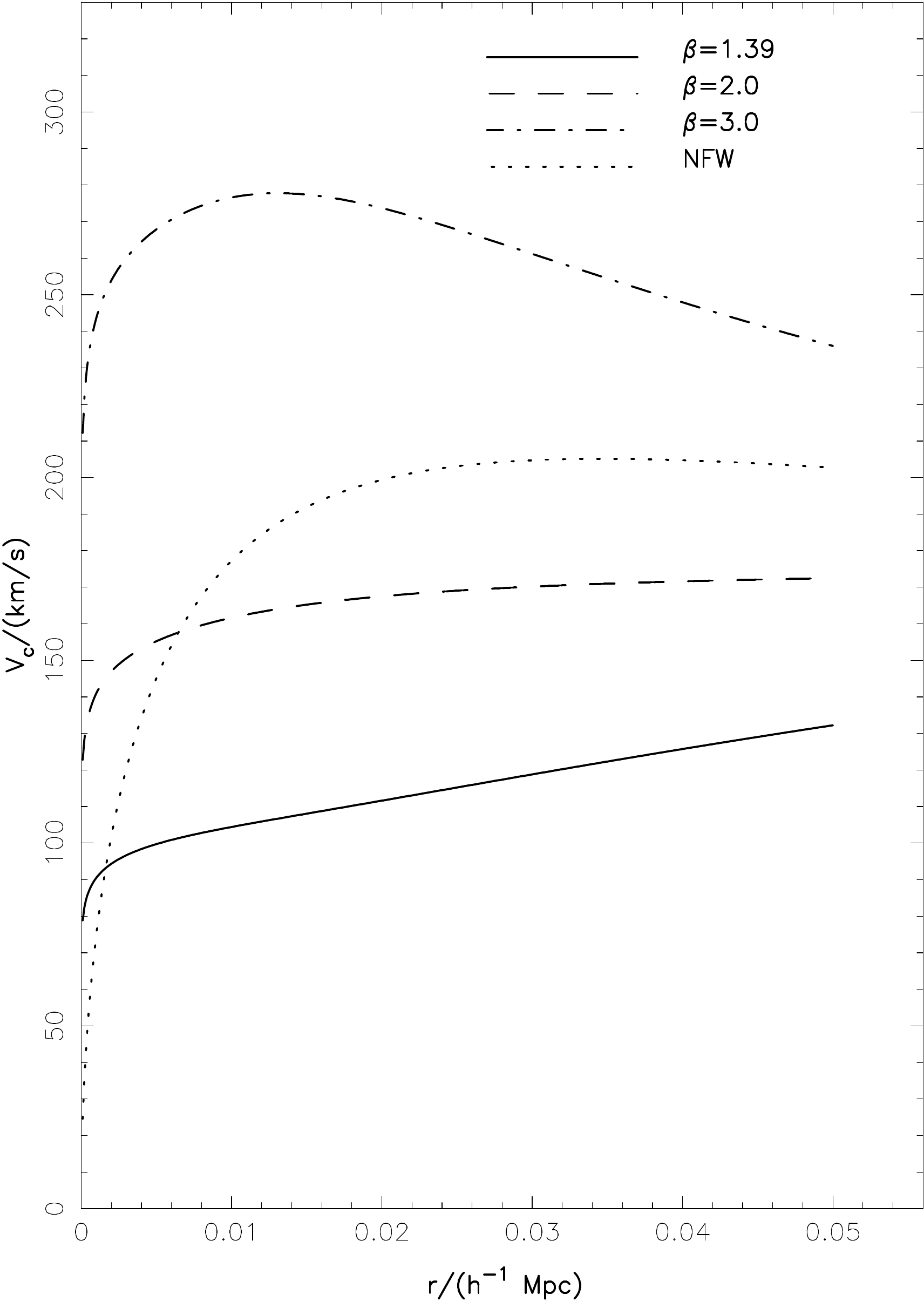}}
}
\caption{ The rotation curves for NFW model (dotted line), our fit model 
$(\alpha,\beta,\gamma)=(3.04,1.39,1.88)$ (solid line), and the cases when
$(\alpha,\beta,\gamma)=(3.04,2.0,1.88)$ (dashed line) and 
$(\alpha,\beta,\gamma)=(3.04,3.0,1.88)$ (dot-dashed line).} 
\label{fig5}
\end{center}
\end{figure}

We have repeatedly emphasized that strong lensing is very sensitive to the 
inner slopes of the assumed density profiles of lensing halos 
\citep{Chen05,LC09,CM10}, however, this is true only when the concentration 
parameters ($c=r_{200}/r_{\rm s}$ for NFW and the like, CP hereafter) for 
different models are fixed in actual lensing probability calculations 
\citep{Sarbu01,li02,Chen03a,Chen03b,Chen04a,Chen04b,Zhang04}. We know that  
CP reflects a slope distribution starting from the 
innermost region out to the outer regions of a halo. Hence if CP is treated as 
a changeable parameter, it would be more important than  
the inner slope  to lensing efficiency \citep{Giocoli12}. On the other hand, 
just like strong lensing efficiencies, different rotation curves are 
predicted by different total (DM+baryon) density profiles of  galactic halos, 
and hence are also very sensitive to the concentration parameter $c$ 
\citep{McGaugh16}. Therefore, both lensing efficiencies and rotation curves are 
sensitive not only to inner slopes but also to outer slopes. For strong 
lensing, the outer slope of our fitted profile ($\beta\sim 1.39$) is smaller 
than Schaller15 profile ($\sim 3$), which results in the overpredictions of the 
latter to the observations of SQLS12 (Fig.\ref{fig1}). For rotation curves, 
this can be easily verified by changing outer slopes. To see this, 
in Fig.\ref{fig5}, we present the rotation curves that are predicted based on 
our fit model (solid line), and the models by keeping 
$\alpha=3.04$ and $\gamma=1.88$ unchanged, while increasing outer slope 
$\beta$ to $2$ (SIS like, dashed line) and $3$ (NFW like, dot-dashed line). 
Cleanly, the outer parts of rotation curves increases with increasing outer 
slope.

\section{Discussions and Conclusions }
\label{sec:conc}
We have used the $(\alpha,\beta,\gamma)$ double power-law 
model to fit the most recent strong lensing observations SQLS12, 
and get the best-fit parameters $\alpha$=3.04, $\beta$=1.39 and 
$\gamma$=1.88. It turns out that the outer part of the rotation curve predicted 
by this fitted density profile is much lower than that predicted by other 
models which can account for the observations. So we conclude that the density 
profile of DM haloes that can predict the most recent observations of strong 
lensing cannot predict the observations of rotation curves.

In a certain sense, we admit that, the one-population, $(\alpha,\beta,\gamma)$ 
double power-law model is a toy model when applied to strong lensing sample 
SQLS12 which include both galaxies and clusters of galaxies. In strong lensing 
statistics, we usually employ a two-population SIS+NFW model. However, as is 
pointed out earlier, in this two-population model, NFW is not enough to account 
for the lensing rates of large image separations. What's more, we indeed have a 
``universal'' density profile, NFW, which still serves us as starting point in 
current N-body cosmological simulations. Including baryon effects can only 
change the slopes near the centers of haloes.  In particular, Schaller15 model 
is only the most recent example of the similar simulations which are of 
one-population and are valid both for galaxies and clusters 
\citep{SIR99,MCW06,MWC08}. Of course, it seems impossible that, in simulations, 
the baryon effects can reproduce such shallow outer slopes as $\beta\sim 1.4$ 
of our fit model. The problem is, even if some mechanisms in simulations can do 
the job, this shallow outer slope is incompatible with the observations of 
rotation curves.

At first sight, it is a temptation to fit the lensing sample with 
$(\alpha,\beta,\gamma)$ double power-law model but allow 
$\alpha$,$\beta$, and $\gamma$ all to be halo mass-dependent, as done by 
\citet{Cintio14} when they fit their simulations. Unfortunately, more than a 
dozen of free parameters involved and the high degeneracy between lensing 
probabilities and density profiles make it impossible for merit functions to be 
convergent. 

On the other hand, we may consider some two-population fittings. As is 
well-known, SIS has been proved to be a good model to account for giant 
ellipticals as lenses. But SIS is in conflict with the observations of 
rotation curves which indicates a constant core for most disk galaxies 
\citep{deBlok01,deBlok08,deBlok10,McGaugh07}. As for clusters of galaxies, they 
have nothing to do with rotation curves. So two-population fittings cannot do 
better than  one-population $(\alpha,\beta,\gamma)$ double power-law model on 
the subject in this paper.

As a summary, in paper I and other previous works, we find that the density 
profiles of DM haloes which are in agreement with observations of rotation 
curves cannot match the observations of strong lensing probabilities; in this 
paper, we find that the converse is also true, namely, the density profiles 
fitted from strong lensing probabilities cannot predict the observations of 
rotation curves. So we conclude that, in $\Lambda$CDM paradigm, it is difficult 
to reconcile the contradictions between the observations of rotation curves 
and strong gravitational lensing.

\section*{Acknowledgements}
 This work was supported by the National Natural Science Foundation of China 
(Grant 11073023).

\setlength{\bibhang}{2.0em}

\end{document}